\documentclass[]{article}
\usepackage{times,graphicx,a4wide, natbib}

\pdfoutput=1
\begin{document}
	
\title{The Galactic Club, or Galactic Cliques? Exploring the limits of interstellar hegemony and the Zoo Hypothesis}
\author{Duncan H. Forgan$^1$}
\maketitle

\begin{center}

$^1$Scottish Universities Physics Alliance (SUPA),\\School of Physics and Astronomy, University of St Andrews \\

\end{center}

\noindent \textbf{Word Count: 3,405} \\

\noindent \textbf{Direct Correspondence to:} \\
D.H. Forgan \\
\textbf{Email:} dhf3@st-andrews.ac.uk \\

\newpage

\begin{abstract}

\noindent The Zoo solution to Fermi's Paradox proposes that extraterrestrial intelligences (ETIs) have agreed to not contact the Earth.  The strength of this solution depends on the ability for ETIs to come to agreement, and establish/police treaties as part of a so-called ``Galactic Club''.  These activities are principally limited by the causal connectivity of a civilisation to its neighbours at its inception, i.e. whether it comes to prominence being aware of other ETIs and any treaties or agreements in place.

If even one civilisation is not causally connected to the other members of a treaty, then they are free to operate beyond it and contact the Earth if wished, which makes the Zoo solution ``soft''.  We should therefore consider how likely this scenario is, as this will give us a sense of the Zoo solution's softness, or general validity.

We implement a simple toy model of ETIs arising in a Galactic Habitable Zone, and calculate the properties of the groups of culturally connected civilisations established therein.  We show that for most choices of civilisation parameters, the number of culturally connected groups is greater than 1, meaning that the Galaxy is composed of multiple Galactic Cliques rather than a single Galactic Club.  We find in our models for a single Galactic Club to establish interstellar hegemony, the number of civilisations must be relatively large, the mean civilisation lifetime must be several millions of years, and the inter-arrival time between civilisations must be a few million years or less.

\end{abstract}


\newpage

\section{Introduction}\label{sec:intro}

Fermi's Paradox remains insoluble to humankind. The lack of observational data for extraterrestrial intelligences (ETIs), known commonly as Fact A \citep{Hart1975}, must be reconciled with our understanding of our own civilisation, which we might assume is not rare or unique thanks to the Copernican Principle of Mediocrity.  \citet{BrinG.D.1983} and \citet{fermi_review} review the Paradox in detail.

One particular solution to the Paradox is referred to as the Zoo hypothesis \citep{Ball1973}. In this scenario, humanity is deliberately kept out of the Galactic conversation for one or more reasons, ranging from our own primitive nature and a desire to protect Earth as a nature reserve, or perhaps a recognition that contacting less developed civilisations has a deleterious effect on their development.  The related Interdict Solution also proscribes ETIs from making contact or revealing themselves to us for both our sake and theirs \citep{Fogg1987}.

However, this solution (and many others like it) demand a uniformity of motive amongst ETIs. An interdict placed on Earth can be utterly broken by a single message or spacecraft. Social norms, especially those cultivated between civilisations that evolved independently of each other, require policing.

The development and policing of social norms requires, at the very least, causal contact between civilisations, and the existence of a ``Galactic Club'' to agree on these norms, as well as jurisprudence to deal with their violation (cf \citealt{Freitas1977}).  \citet{Hair2011} argued that in a simple model of civilisation arrival, if the distribution of individual arrival times is Gaussian, then the time between the appearance of the first and second civilisations in the Milky Way, $IAT_1$, follows an inverse exponential distribution \citep{Snyder1991}. This inter-arrival time can therefore be very large, meaning the first civilisation to arise is able to influence the others greatly, and thereby facilitate the setup of social norms and uniformity of motive across the entire Galaxy.

\citet{Forgan2011b} argued against this model, noting that space-time separation is the critical variable for cultural connectivity, and hence the spatial distribution of civilisations is likely to break this hegemony. Note that this definition of cultural connection demands that a civilisation begins receiving transmissions from other civilisations before it becomes technologically advanced enough to detect them, and hence cannot develop its own customs regarding other civilisations without being influenced by previously established norms.

However, the extent of \citet{Forgan2011b}'s work was to suggest that for a plausible set of civilisation properties, the number of culturally-connected civilisation groups (CCGs) in the Milky Way $N_{group} >1$.  The next logical step is to investigate the behaviour of $N_{group}$ as a function of the properties of ETIs, and identify regimes where a single Galactic Club might be established, and where multiple, smaller Galactic Cliques are established. 

In this work, we investigate a toy model for the emergence of intelligent civilisations in the Milky Way.  By measuring the space-time separations of civilisation pairs, we establish groups of civilisations that are culturally connected. By doing this we can investigate the conditions required for the establishment of uniformity of motive amongst a Galactic population of intelligent species.

In section \ref{sec:method} we describe the simulation techniques we adopt to model the causal and cultural connections between civilisations; section \ref{sec:results} describes the resulting groups established, and the likelihood that the Galaxy contains a single Club, or multiple Cliques; and we summarise the work in section \ref{sec:conclusion}.

\section{Method} \label{sec:method}

We carry out Monte Carlo Realisation simulations of civilisation emergence (and extinction), using a simple toy model.   We place civilisations in a Galactic Habitable Zone similar to that of \citet{GHZ} and \citet{Gowanlock2011}.  The field of Galactic Habitability is struggling to achieve consensus on the true GHZ, and it is clear that it will depend sensitively on the stellar kinematics as well as the hierarchical merging history of the Milky Way \citep{Forgan2015,Vukotic2016}. We will return to the assumptions made regarding spatial distribution in the Discussion. 

Our GHZ extends from 6kpc to 10kpc, with an exponentially decreasing surface density of stars with radius:

\begin{equation}
\Sigma_{*}(r) \propto e^{-R/R_s}.
\end{equation}

\noindent The scale length $R_s= 3 $kpc.  For simplicity, we do not model the vertical stratification of the Galactic disc, and assume that stars are evenly distributed in the z axis between -1 and 1 kpc.

Civilisations are assigned an arrival time, which is sampled from a Gaussian distribution with mean and variance $(\mu_{arrive}$, $\sigma^2_{arrive})$.  This parametrisation reflects the observation that the factors which govern the emergence of a civilisation satisfy the conditions for the application of the Central Limit Theorem, which has been demonstrated by more detailed MCR modelling \citep{mcseti1,mcseti2}. Rather than attempt to constrain the parameters, we instead explore a larger parameter space, presumably larger than the space bounded by factors such as the star formation history and age-metallicity relation of the Milky Way \citep{Rocha_Pinto_AMR, Rocha_Pinto_SFH}, and the details of what makes a planet habitable, what keeps it habitable \citep{Raup_and_Sepkoski, Rushby2013,O'Malley2013}, what governs the emergence of life and intelligent life \citep{stages}, and the essentially unknown sociological factors that govern a civilisation's development and lifetime.

To calculate civilisation connectivity, we calculate the space-time separation 4-vector 

\begin{equation}
dx_{\nu} = \left( c \Delta t, \Delta x,\Delta y,\Delta z \right)
\end{equation}

\noindent where $c$ is the speed of light, $\Delta t$ is the difference in arrival time between the two civilisations, and $\Delta x$, $\Delta y$ and $\Delta z$ are the spatial separations in Cartesian co-ordinates.  We adopt the following convention:

\begin{equation}
\left|dx_{\nu}\right|^2 = c^2 \Delta t^2 - (\Delta x^2 + \Delta y^2 + \Delta z^2) \label{eq:4vec}
\end{equation}

\noindent And hence for two civilisations to be causally connected, $\left|dx_{\nu}\right|^2$ must be positive (or equivalently, $\left|dx_{\nu}\right|$ must be real).  Note that this 4-vector represents the strictest constraints on two civilisations being connected and aware.  In effect, it demands that a signal transmitted from civilisation $i$ reaches civilisation $i+1$'s home planet before civilisation $i+1$ emerges.  We could construct similar 4-vectors for other possibilities, such as crewed or uncrewed spacecraft being sent from $i$ to $i+1$.  This would require modification of equation (\ref{eq:4vec}), replacing $c$ with a variable representing the spacecraft velocity, which special relativity demands must be less than $c$.  We can therefore be confident that if the space-time separation as given by equation (\ref{eq:4vec}) is negative, then it must be negative regardless of how a civilisation attempts to communicate.

We assign each civilisation a lifetime, which is also sampled from a Gaussian defined by its mean and variance $(\mu_{life},\sigma^2_{life})$.  We therefore demand that a communication window between both civilisations is open, i.e. that both civilisations must be able to communicate before one or the other goes extinct.

We calculate connected groups using the following algorithm:

\begin{enumerate}
\item  Firstly, the set of all civilisations is sorted by arrival time.  The first civilisation to arrive establishes the first group, and it is identified as that group's ``leader''.  
\item We then test all other civilisations against the leader, in ascending arrival time order, for causal connections using equation (\ref{eq:4vec}).  
\item If the space-time separation between the leader and a civilisation is positive, the civilisation joins the leader's group.
\item If a civilisation is not connected to the leader, it begins its own group and is established as a leader.  
\item Once all civilisations are tested, we move to the next civilisation that is not connected, and repeat the algorithm until all civilisations belong to a group.
\end{enumerate}

This produces a single realisation of the culturally connected groups in the GHZ.  We carry out 30 realisations for any given set of parameters $(N_{civ},\mu_{arrive}, \sigma_{arrive},\mu_{life},\sigma_{life})$, and use this data to compute mean and standard deviations on the resulting statistics, which include the number of groups $N_{group}$ and their maximum extent $S_{group}$, which is defined as the maximum distance between any two pairs of civilisations in the group.

\section{Results \& Discussion}\label{sec:results}

\subsection{Dependence on Total Number of Civilisations}

\begin{figure}
\begin{center}
\includegraphics[scale=0.5]{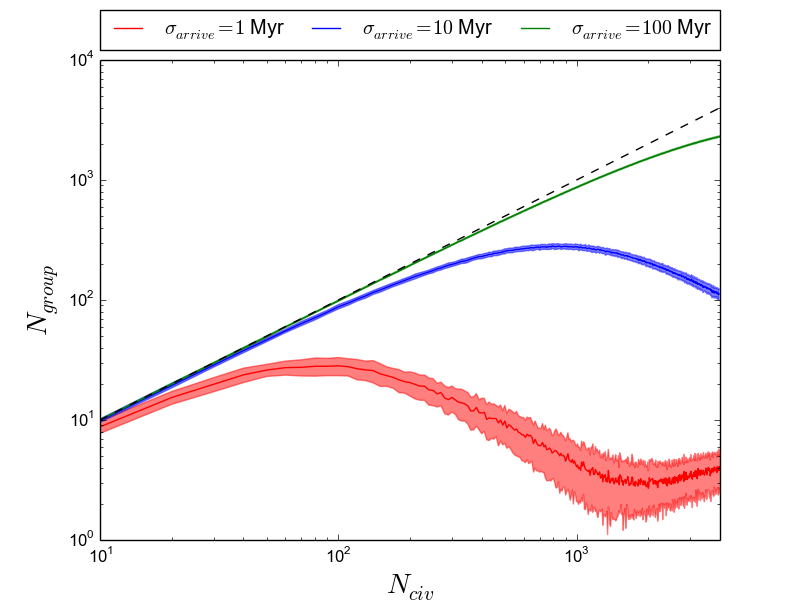}
\caption{The mean number of groups identified in the MCR runs as a function of civilisation number, for three values of $\sigma_{arrive}=1,10,100$ Myr.  The black dashed line indicates the maximum number of groups $N_{group}=N_{civ}$.  We fix $(\mu_{life},\sigma_{life}) = (0.1, 10^{-3})$ Myr and $\mu_{arrive}=5000$ Myr.}
\label{fig:ngroup_vs_nciv}
\end{center}
\end{figure}

\noindent We first begin by fixing all parameters, and carry out a series of realisations for multiple values of $N_{civ}$.  Figure \ref{fig:ngroup_vs_nciv} shows how the mean number of causally-connected groups (CCGs) varies with increasing $N_{civ}$ for a fixed $(\mu_{life},\sigma_{life}) = (0.1, 10^{-3})$ Myr and $\mu_{arrive}=5000$ Myr.  Each curve represents a different value of $\sigma_{arrive}=(1,10,100)$ Myr.

\begin{figure}
\begin{center}
\includegraphics[scale=0.5]{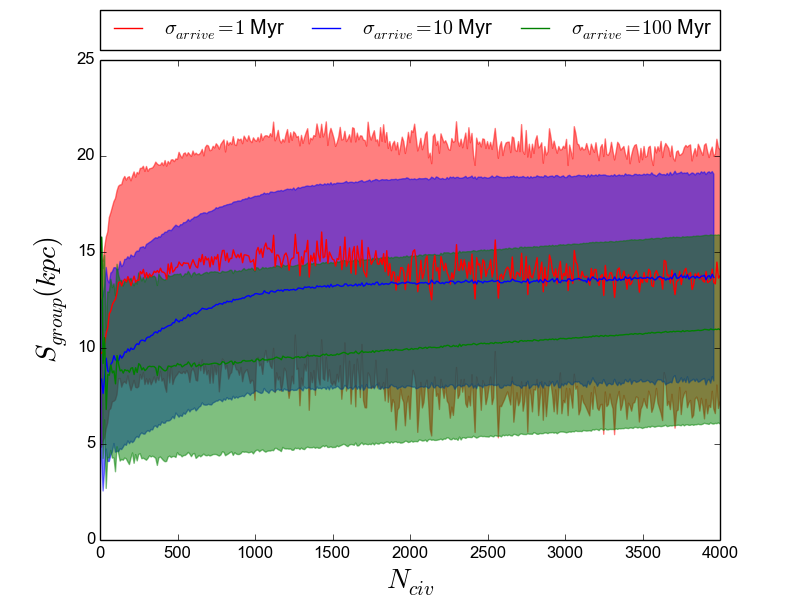}
\caption{The mean spatial extent of groups identified in the MCR runs as a function of civilisation number, for three values of $\sigma_{arrive}=1,10,100$ Myr. As in Figure \ref{fig:ngroup_vs_nciv}, $(\mu_{life},\sigma_{life}) = (0.1, 10^{-3})$ Myr and $\mu_{arrive}=5000$ Myr.}
\label{fig:groupsize_vs_nciv}
\end{center}
\end{figure}

We can immediately see that the lowest group counts occur when $\sigma_{arrive}$ is at its minimum value.  This is at direct odds with the result of \citet{Hair2011}, which prefers a relatively large value of $\sigma_{arrive}$ for hegemony establishment based on the consequently large inter-arrival time between the first and second civilisations.  However, we find that for the largest values of $\sigma_{arrive}$, the number of groups $N_{group}$ is at its maximum (i.e. it is equal to the number of civilisations $N_{civ}$) until it reaches $N_{civ}>300$.  In all three cases, the mean spatial extent of each civilisation asymptotes to similar values at large $N_{civ}$ (Figure \ref{fig:groupsize_vs_nciv}).  While the $\sigma_{arrive}=100$ Myr sizes are slightly lower than the other two cases, it remains within a single standard deviation.  

In the case where $\sigma_{arrive}$ is set to its lowest value, the minimum number of groups tends to around 3, and asymptotes to this value for $N_{civ}>1000$ (where it also approaches the maximum group extent of 20 kpc, corresponding to the diameter of the Galactic Habitable Zone edge).  Given that the $1\sigma$ uncertainty on $N_{group}$ is around 1, this result contains the Galactic Club scenario within the 3$\sigma$ confidence interval.  We can therefore predict that for this Galactic Habitable Zone configuration, civilisation populations emerging within a relatively narrow time interval have the best odds for establishing the Galactic Club (given a typical civilisation lifetime of 0.1 Myr).  

Such co-ordination of arrival time seems a priori unlikely - indeed, it most likely demands that a global regulation mechanism exists to ``synchronise the biological clocks'' of planets separated by enormous distances \citep{Vukotic_and_Cirkovic_07,Annis}.  Proposed regulation mechanisms, such as gamma ray bursts, would still fail to synchronise the entire GHZ, as their ability to sterilise planets extends no more than a few kpc, and their efficacy even at these distances remains a source of debate \citep[][and references within]{Martin2009,Thomas2009}.

\subsection{Dependence on Civilisation Lifetime Parameters}

\begin{figure*}
\begin{center}$\begin{array}{c}
\includegraphics[scale=0.4]{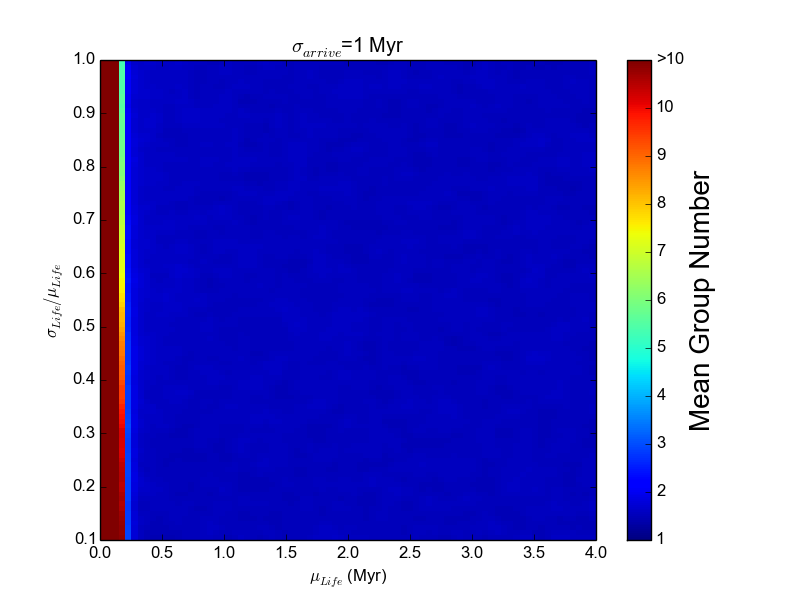} \\
\includegraphics[scale=0.4]{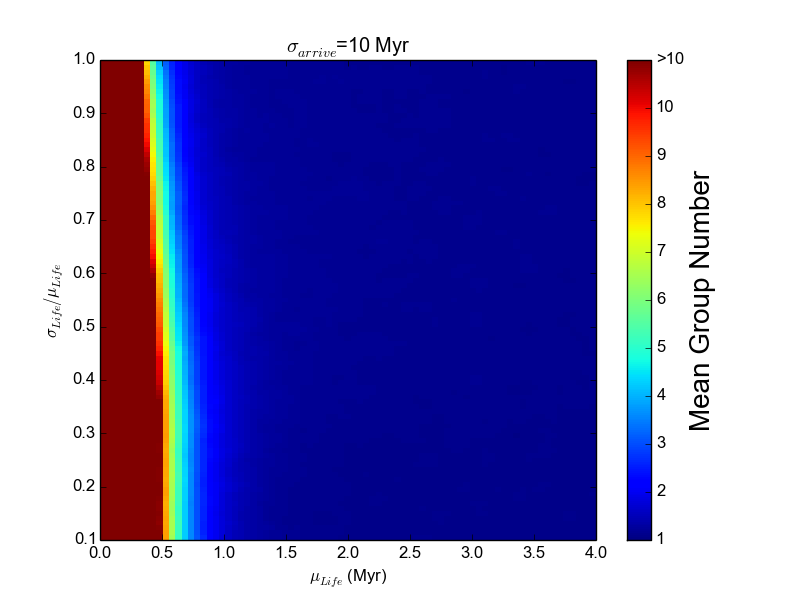} \\
\includegraphics[scale=0.4]{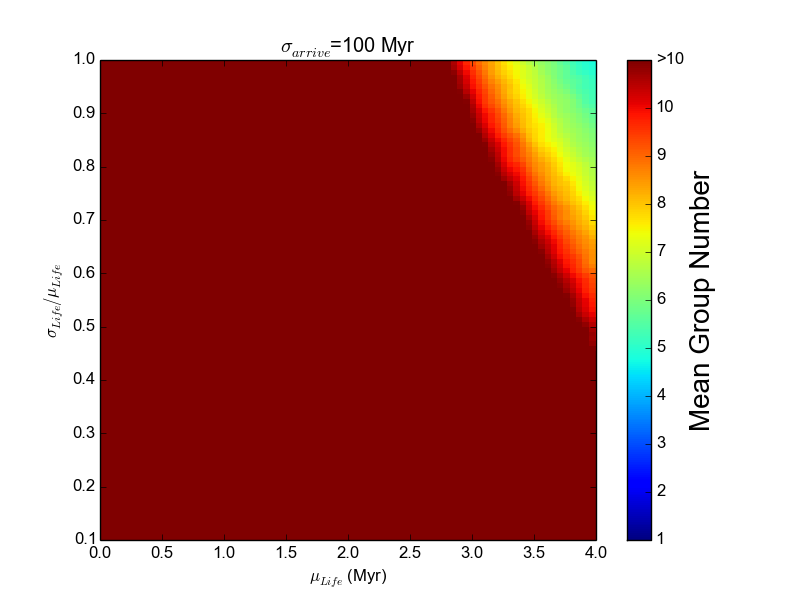} \\
\end{array}$
\caption{The mean number of groups as a function of $\mu_{life}$ and $\sigma_{life}$.  The mean arrival time $\mu_{arrive}$ is held fixed, with each plot denoting a different value of $\sigma_{arrive}$.   $\mu_{arrive}$ is fixed at 5000 Myr.}
\label{fig:mu_vs_sigma}
\end{center}
\end{figure*}

As we have established that $N_{group}$ can become small for $N_{civ}>500$, we now fix $N_{civ}=500$ and explore the effects of the civilisation lifetime parameters $(\mu_{life}, \sigma_{life})$.  Figure \ref{fig:mu_vs_sigma} shows how the mean group number depends on these parameters (with each plot showing a different value of $\sigma_{arrive}$).  Again, we can see that hegemony establishment ($N_{group}=1$) is easier if $\sigma_{arrive}$ is lower.  The group number is only weakly dependent on $\sigma_{life}$ (except for very high $\sigma_{arrive}$).

In all cases, the mean civilisation lifetime $\mu_{life}$ must exceed around 250,000 years for hegemony to be established.  Note that this lifetime is measured from when a civilisation is sufficiently technologically advanced to begin communication.  The earliest fossil records of \emph{homo sapiens} date to approximately 190,000 years ago \citep{McDougall2005}.  If human civilisation is indeed subject to the Principle of Mediocrity, then we should expect our total lifetime (from the emergence of anatomically modern humans to now, and from now until the end of our civilisation) to be close to the mean.  The results of this toy model suggests that for hegemony establishment to proceed and form a Galactic Club, our species has likely only persisted for approximately half its lifespan.

Note that if we wished a single civilisation to be exceptionally long-lived near the beginning of the simulation (a low mean lifetime with a high standard deviation), then this would still result in a reasonably large number of groups.  This underlines that space-time separation is the principal factor, and that the Galaxy is sufficiently large that cliques can be established at large spatial separations even from ancient long-lived civilisations.

\section{Discussion}

\noindent We should be clear that this work makes no predictions on the number of intelligent civilisations in the Milky Way.  It cannot even predict if extraterrestrial intelligences exist at all.  It is merely a controlled numerical experiment that tests the ability of civilisations to influence each other if they exist, depending on the details of when and where civilisations might emerge.

We have progressed beyond the original assertion of \citet{Forgan2011b}, that typically the number of culturally connected groups is greater than 1.  Our Galactic Habitable Zone model of civilisation emergence predicts that \emph{initially}, there will be significant opportunity for cultural variance across space and time, and as such uniformity of motive is not present.  

To reach this conclusion, we have assumed an annular GHZ for civilisation emergence.  We have assumed the rather wide annular range of 6-10 kpc based on \citet{Gowanlock2011}, which remains the most high-resolution study of galactic habitability, if somewhat constrained by demanding azimuthal symmetry \citep{Forgan2015}.  This is actually a restricted GHZ - Gowanlock found that habitable planetary systems were possible from as little as 2.5 kpc from the Galactic Centre.  Populating this region with civilisations will result in much smaller spatial separations due to the high surface density of stars, which may reduce the number of cliques.  However, the maximum spatial separation is a function of the outer radius of the GHZ:

\begin{equation}
dx_{max} = 2 R_{outer}.
\end{equation}

\noindent If $N_{civ}$ remains small, the outer edge will be poorly populated, reducing the maximum separation below this value (we can infer this by studying the mean group size curves in Figure \ref{fig:groupsize_vs_nciv} for $N_{civ}<100$).  This would increase the probability of a single Club forming, but only in this limit.   If $N_{civ}$ is sufficiently large, populating the interior with more civilisations cannot allow a Galactic Club to form, as a small fraction will reside at the distant fringes of 10 kpc, and the number of culturally connected groups will in general be greater than 1.  This is true as long as $R_{outer}$ is relatively large.  Given that we exist at $R \approx 8$ kpc, we can be reasonably confident that this is the case.

We also assumed that stellar motions are negligible in this analysis, which is clearly not the case even in a relatively restricted GHZ \citep{Vukotic2016}.  While the speed of light \emph{in vacuo} remains constant in all reference frames, the distance between stars can be reduced by proper motions, resulting in reduced light travel times and greater probability of civilisation connectivity (with the converse being true for stars receding from each other).  Once cliques have been established, the spatial extent of the clique will evolve according to local stellar dynamics.  It is likely that initially separate cliques will be able to diffuse into each other.  Given that cliques are somewhat intermixed even at inception (which we can see from the large group extents established in Figure \ref{fig:groupsize_vs_nciv}), it is unclear what further effects this may cause.

While cultural variation is initially large, this does not preclude the later emergence of uniform motive once individual cliques become causally connected to each other, i.e. a uniformity established through political means.  The galactopolitical machinations of a set of civilisation cliques (and the internecine activities of an individual clique) far exceed the capabilities of this simple toy model. All that we can predict is that if civilisation cliques do come into contact, it is likely they will hold significantly different perspectives on the Universe, and the rights and responsibilities of sentient beings and the institutions they construct.

It is also possible for cliques to evolve internally, in isolation from their peers.  Cultural norms change with time, and the growth of a clique as new civilisations become culturally connected may enhance the rate of cultural evolution, as new ideas and perspectives begin to percolate through the clique's membership.  A clique that initially holds treaties regarding contact in high regard may discard them through changes in their internal make-up, especially if they are subject to strong environmental pressures that impact their way of life.

Despite these factors that are beyond the realms of simple MCR analyses such as this work, we are still able to draw important conclusions on the Zoo Hypothesis.  The initial state of the Zoo hypothesis (that is, when cliques initially come into being) is soft - in general, we must assume the Galaxy is culturally diverse.  Subsequent cultural evolution can act to soften the Zoo Hypothesis, just as much as it may act to harden it.  For example, a dominant clique may attempt to impose an authoritarian monoculture, or cliques may ``agree to disagree'' for political expediency.  

We have no way of predicting how extraterrestrial cultures will interact.  However, we do know that from the multitude of possible outcomes of political negotations, the Zoo hypothesis demands that only a small subset of these outcomes are possible.  If the Zoo hypothesis is correct, and it demands a uniformity of motive established via a Galactic monoculture, we should conclude that it is most likely imposed  - perhaps against the wishes or interests of the Galactic community - through interactions between a number of cliques, either through political or military means.

% GHZ configuration
% Cultural evolution of individual cliques
%

\section{Conclusion}\label{sec:conclusion}

\noindent In this work, we have investigated the implicit assumptions made to invoke the Zoo hypothesis as a solution to Fermi's Paradox.  We achieved this by explored the culturally-connected civilisation groups present in a toy model of the intelligent civilisation population of the Milky Way.  We find that for there to be only a single group (a ``Galactic Club''), the mean civilisation lifetime must be extremely long, and the arrival time between civilisations must in fact be relatively short (constrained by a small standard deviation in arrival time, $\sigma_{arrive}$).  This is perhaps an unlikely scenario, as it would require a large number of civilisations to emerge across the Galaxy in a very short time frame.  This is also in opposition to previous work which has suggested $\sigma_{arrive}$ needs to be large for a single first civilisation to dominate the Milky Way.

This toy model underlines that the Zoo solution to the Fermi Paradox remains ``soft'', i.e. it demands a uniformity of motive amongst ETIs that only exists when certain conditions are met.  Our previous work in this area has suggested that these conditions are unlikely to be met in our Galaxy \citep{Forgan2011b}, but did not explore the population of causally-connected groups that would result.

We therefore conclude the following:

\begin{itemize}
\item If civilisations typically last less than 1 Myr, then it appears likely that $N_{group}>>1$, resulting in a set of ``Galactic Cliques'' rather than a single Galactic Club. 
\item A single long-lived, ancient civilisation still fails to knit the entire Galactic community of civilisations into a single Club. 
\item If \emph{all} civilisations can last much longer than 1 Myr, then a single Galactic Club can be established, but only if all civilisations arrive quite close together in time.
\end{itemize}

Typically, the Galaxy is composed of ``Galactic Cliques''.  Once established, these cliques may come into causal contact with others, bringing their own established norms to discussions in a scaled-up version of contact between individual civilisations.  One clique attempting to place an interdict on contacting ``primitive'' civilisations is likely to encounter significant problems if another clique disagrees.

This analysis remains insufficient to completely remove the Zoo solution from the list of possible solutions to Fermi's Paradox, but it illustrates the underlying assumptions required to propose it.  It may well still be the case that the Earth resides in a region of space occupied by a conservative clique bent on non-contact.  However, as our ability to detect unintentional signals from both living and dead civilisations increases (e.g. \citealt{Wright2015,Stevens2015}), we should presumably be able to break the deadlock imposed in this scenario.  In an extreme case, a neighbouring clique is free to violate non-contact treaties with impunity - if we are a late arrival to a populous Galactic community, then many established cliques may be aware of our presence, and they may not pay attention to local signage forbidding them to reveal themselves.

\section*{Acknowledgments}

The author gratefully acknowledges support from the ECOGAL project, grant agreement 291227, funded by the European Research Council under ERC-2011-ADG, and the STFC grant ST/J001422/1. 

\bibliographystyle{mn2e} % (must include a bibliography style)
\bibliography{galactic_cliques}

\end{document}